\documentstyle[11pt]{article}
\renewcommand{\thefootnote}{\fnsymbol{footnote}}

\def\appendix#1{
  \addtocounter{section}{1}
  \setcounter{equation}{0}
  \renewcommand{\thesection}{\Alph{section}}
 \section*{Appendix \thesection\protect\indent \parbox[t]{11.715cm} {#1}}
  \addcontentsline{toc}{section}{Appendix \thesection\ \ \ #1}
  }

\def\be{\begin{equation}}
\def\ee{\end{equation}}
\def\beq{\begin{equation}}
\def\eeq{\end{equation}}
\def\bea{\begin{eqnarray}}
\def\eea{\end{eqnarray}}

\begin{document}

\begin{titlepage}
\begin{flushright}
NBI--HE--97--11\\
hep-th/9703208\\
\end{flushright}
\vspace{.5cm}

\begin{center}
{\LARGE A note on $T^5/Z_{2}$ compactification of the M-theory matrix
model}\\
\vspace{1.2cm}
{\large Ansar Fayyazuddin\footnote{E-mail: ansar@nbi.dk},
Douglas J. Smith\footnote{E-mail: D.Smith@nbi.dk}}\\
\vspace{24pt}
{\it The Niels Bohr Institute,}
\\ {\it  Blegdamsvej 17, DK 2100 Copenhagen \O, Denmark}
\end{center}
\vskip 0.9 cm
\begin{abstract}
We study the $T^{5}/Z_2$ orbifold compactification of the M-theory
matrix model.  This model was originally studied by Dasgupta, Mukhi, 
and Witten.
It was found that one had to add 16 $5$-branes to the system to make the
compactification consistent.  We demonstrate how this is mimicked
in the matrix model.  
\end{abstract}

\end{titlepage}
\setcounter{page}{1}
\renewcommand{\thefootnote}{\arabic{footnote}}
\setcounter{footnote}{0}

A recent proposal by Banks, Fischler, Shenker and Susskind (BFSS) \cite{bfss}
that M-theory in an infinite momentum frame is described by a certain 
matrix model has excited much attention and activity.  This proposal has passed
a number of non-trivial tests.  There are indications, however, that
there are certain elements still lacking in this formulation.  By putting
the matrix model to various tests one hopes to understand its strengths
and weaknesses leading ultimately to, hopefully, the correct formulation
of M-theory.
	The purpose of this note is to put the matrix model to another
test which, in our opinion, the matrix model passes successfully.  
We consider compactification of the matrix theory on a five dimensional
torus modded out by a $Z_2$ action which is supposed to be the analogue
of a compactification considered by Dasgupta, Mukhi and Witten
\cite{dm, wit5}.  This compactification
is rather involved due to constraints coming from gravitational anomaly
cancellation and cancellation of total charge on the compact manifold.
It is of some interest then
to see how these constraints arise in the matrix model compactification.

Compactification on a $d-$dimensional torus, according to the
rules and explicit constructions of references \cite{wtaylor, ls,
ganoretal, lstorus}, results in supersymmetric Yang-Mills theory in
$d+1$ dimensions (SYM$_{d+1}$)
with $16$ super-charges.  For example, compactification on $T^3$ 
results in ${\cal N} = 4$ supersymmetric Yang-Mills theory in $d=4$
\cite{ls, ganoretal}.
Toroidal compactifications always result in non-chiral gauge theories.
As in ordinary string theory compactifications, one can get chiral
theories by considering orbifolds of tori.  

The BFSS model is ${\cal N } = 1$ supersymmetric Yang-Mills in $10$ dimensions
with gauge group SU($N$) dimensionally reduced to $d=1$.  The
resulting theory is matrix quantum mechanics of $9$ bosonic 
and $16$ fermionic matrices
transforming in the adjoint of the SU($N$) gauge group. The limit
$N \rightarrow \infty$ is to be taken. This theory
has only $16$ supercharges 
which is the appropriate number of manifest supersymmetries in the
infinite momentum frame for a theory with $32$ supercharges.
The Lagrangian for the theory is given by:
\begin{equation}
{\cal L} = {\rm Tr} \left ( \frac{1}{2R}(D_0X^a)^2 +
	\frac{R}{4}[X^a,X^b]^2 + \overline{\Psi}D_0\Psi +
	iR\overline{\Psi}\Gamma_a[X^a,\Psi] \right ) \label{Lagrangian}
\end{equation}
where $X^a$ ($a=1, \ldots , 9$) and $\Psi$ are $N \times N$ matrices. Here
the $\Gamma_a$ are $32 \times 32$ 10-dimensional gamma matrices. $\Psi$ is
real and satisfies $\Gamma_{11}\Psi=\Psi$.

When one compactifies direction $i$ on $S^1$ one essentially
replaces $X_{i}$ by a covariant derivative in the $i$ direction.  
This rule is the reverse of the dimensional reduction procedure
where one replaces a covariant derivative by the gauge field which
appears as a scalar in the reduced theory with no dependence on the 
compactified coordinate.  This logic then tells
us that the compactification of the matrix model on $T^5$ is the
same as the dimensional reduction of ${\cal N } = 1$ SYM$_{10}$ to $d=6$.
The resulting theory is then ${\cal N} = (8,8)$
\footnote{We shall use this notation to specify the number of real
supercharges of each chirality.}
super Yang-Mills in $d=6$ \cite{lstorus}.

Orbifolds of matrix models have been considered recently in 
\cite{ulfgab, ks, motl, kimrey, lowe}. 
We are interested in a $Z_2$ orbifold of the above toroidal compactification
where one reverses the
orientation of the compactified coordinates.  This is not a symmetry
of the theory so one has to combine this with another action on the
fields such that the combined transformation is a symmetry of the theory.
We will denote the orientation reversal of the compactified directions
by $R$ and the second transformation by $S$. We will describe the physical
interpretation of these transformations for the bosonic variables. The
interpretation of the transformation of the fermionic matrices is not so
clear. However, similar transformations are required for invariance of
the matrix model Lagrangian, eq.~(\ref{Lagrangian}). We will follow 
\cite{ulfgab} and \cite{kimrey} in doubling the size of all the matrices
to $2N\times 2N$ in an attempt to describe the states in the theory along
with their images.  All matrices can then be expressed in the block form:  
\beq
X = \left(\begin{array}{ll}
             X_{11} & X_{12} \\ X_{21} & X_{22}
          \end{array} \right).
\eeq
Using the notation of \cite{kimrey} we denote by $X_{\bot}$ the compactified
coordinates and by $X_{\|}$ the uncompactified ones. So $X_{\bot}$
represents any of the matrices $X_1, \ldots, X_5$ and similarly $X_{\|}$
represents any of the matrices $X_6, \ldots, X_9$. The action
of $R$ should exchange the blocks $1 \leftrightarrow 2$ and multiply
the $X_{\bot}$ matrices by a minus sign, leaving the rest invariant
\cite{kimrey}:
\bea
R: X_{\bot} &\rightarrow &-MX_{\bot}M \nonumber \\
   X_{\|} &\rightarrow & MX_{\|}M \\
   \Psi &\rightarrow & \gamma_7 M\Psi M \;\; , (\gamma_7 \equiv 
\Gamma^0\Gamma^{1}\cdots\Gamma^{5})\nonumber.
\eea
where $M$ interchanges the blocks.  In \cite{kimrey} two different
matrices $M$ are considered: $M_{1}=\sigma_{1}\otimes 1_{N \times N}$ and
$M_{2}= \sigma_{2}\otimes 1_{N \times N}$.

However, as we have stated earlier, $R$ is not a symmetry of the matrix model.
We must consider another transformation $S$ which, along with $R$,
is a symmetry. The reason $R$ is not a symmetry can be
understood by considering the analogous transformation in M-theory (or more
precisely 11-dimensional supergravity). There $R$ corresponds to a parity
transformation. It is well known that parity alone is not a symmetry due to
the presence of the Chern-Simons term. Therefore it
must be combined with the transformation of the 3-form gauge potential:
\begin{equation}
A \rightarrow -A. 
\label{A-A}
\end{equation}
The question is to find the corresponding transformation
$S$ in the matrix model. One way to do this is to note that the transformation
eq.~(\ref{A-A}) reverses the charge of a membrane. So we should look for a
transformation $S$ that changes the sign of the membrane charge in the matrix
model. In \cite{bss} the supersymmetry algebra and central charges are
calculated explicitly. The charge for a membrane is:
\begin{equation}
Z^{ab} = \frac{i}{2}{\rm Tr}[X^a,X^b]
\end{equation}
If we choose the transformation $S$ to be:
\begin{eqnarray}
X_{\bot} & \rightarrow & X_{\bot}^{\rm T} \\
X_{\|} & \rightarrow & X_{\|}^{\rm T} \\
\Psi & \rightarrow & \Psi^{\rm T}
\end{eqnarray}
it is easy to see that:
\begin{equation}
Z^{ab} \rightarrow -Z^{ab}.
\end{equation}

So we find, as in \cite{kimrey}, that the complete transformation is:
\begin{eqnarray}
X_{\bot} & \rightarrow & -MX_{\bot}^{\rm T}M \nonumber\\
X_{\|} & \rightarrow & MX_{\|}^{\rm T}M \\
\Psi &\rightarrow & \gamma_7 M\Psi^{\rm T}M\nonumber
\end{eqnarray}
It can easily be checked that this combined transformation is a symmetry of
the matrix model Lagrangian, eq.~(\ref{Lagrangian}) in the gauge $A_0=0$.
In the gauge unfixed theory we must also include the transformation:
\begin{equation}
A_0 \rightarrow -MA_0^{\rm T}M
\end{equation}

Toroidal compactification on $T^5$ then leads to $6-$dimensional 
super Yang-Mills theory with ${\cal N}=(8,8)$ supersymmetry with gauge group
SU(2$N$).  The $X_{\bot}$ turn into gauge fields while the remaining
$X_{\|}$ are scalars from the $6-$dimensional field theory perspective.
The fermion field $\Psi$ splits into two opposite chirality complex
Weyl fermions to form a Dirac fermion.  All fields transform in the 
adjoint of the gauge group.  

We can consider the action of identifying under $RS$ on the field theory.  
There are only 2 distinct choices of the matrix $M$ in the above
transformation. As shown in
\cite{kimrey} the choice of $M_1$ leads to a matrix model with gauge group
SO(2$N$) with the negative chirality fermion
transforming in the adjoint
representation and the positive chirality fermion along with $X_{\|}$
transforming in the second rank symmetric 
tensor representation.  
The choice $M_2$ leads to the gauge group USp(2$N$) 
with the negative chirality fermion transforming in the adjoint
representation and the positive chirality fermion along with
$X_{\|}$  transforming in the second rank anti-symmetric 
tensor representation.
These models have ${\cal N}=(8,0)$ supersymmetry.  The negative chirality
fermion combines with the gauge fields (formerly $X_{\bot}$) 
to fill out a vector multiplet
while the positive chirality fermion combines with the four scalars
$X_{6,7,8,9}$ to form a hypermultiplet. 

So compactification of M-theory has led to a six dimensional 
chiral gauge theory with a matter hypermultiplet.  These 
theories have to satisfy stringent anomaly cancellation conditions
as explained in \cite{erler, schwarz, seiberg,intriligator, gabriele}.
The anomaly eight form for six dimensional gauge theories is
\beq
I_8 = {\rm tr}_{adj}F^4 - \sum_{R} n_{R}{\rm tr}_{R}F^4 = \alpha
{\rm tr}_{f} F^4 + c({\rm tr}_{f} F^2)^2. 
\eeq
In \cite{seiberg} it was noted that the gauge theory can be made anomaly
free without introducing gravity if $\alpha =0$ and $c>0$ by the introduction
of a tensor multiplet.  If $c=0$ then the theory is anomaly free
as it stands and no additional multiplets need to be added. 
In the theory at hand we are considering two
cases:

Case 1: G = SO($2N$), with one hypermultiplet in the 
second rank symmetric representation.  In this case the anomaly
is: 
\beq
I_{8} = -16{\rm tr}_{f}F^4.
\eeq
As noted in \cite{gabriele} the addition of hypermultiplets in
other representations cannot set $\alpha =0$.  Thus this theory
cannot be cured of anomalies by the addition of more matter or tensor 
multiplets.

Case 2: G = USp($2N$) with one hypermultiplet in the second rank anti-symmetric
representation.  The anomaly in this case is:
\beq
I_{8} = 16{\rm tr}_{f} F^4. 
\eeq
The anomaly can be cancelled completely in this case by the addition
of $16$ hypermultiplets in the fundamental representation (there is
no need to add tensor multiplets). This has been recently noted in
\cite{gabriele}.

So we are forced to abandon Case 1, and adopt Case 2 to have a sensible
theory.  This case with the addition of $16$ hypermultiplets
in the vector representation has a very nice interpretation due to
Berkooz and Douglas\cite{bd}.  They modelled longitudinal M-theory five-branes
in the matrix model as hypermultiplets in the fundamental 
representation.  Thus we see that to make the matrix model compactification
sensible we are forced to include $16$ five-branes.  This is in complete
agreement with the compactification considered in \cite{wit5} where
one was forced to include $16$ five-branes to cancel the gravitational
anomaly present from the untwisted sector surviving the compactification.

To conclude, the M-theory matrix model has passed a non-trivial test.
The compactification of the M-theory matrix model on $T^{5}/Z_2$
results in an anomalous gauge theory which can be fixed by the addition
of hypermultiplets.  These hypermultiplets are nothing but the Berkooz-Douglas
five-branes \cite{bd} in the matrix model.  This parallels
the low-energy considerations of \cite{wit5}.  We can, conversely, 
view this as evidence that the Berkooz-Douglas five-brane really
is the longitudinal five-brane of M-theory.

\end{document}